\documentclass[reprint,showpacs,preprintnumbers,amsmath,amssymb,superscriptaddress, aps, pra, nofootinbib]{revtex4-1}
\eprint
\usepackage{lipsum}
\usepackage[symbol]{footmisc}
\usepackage{graphicx}
\usepackage{tabularx}
\usepackage{bm}
\usepackage{color}
\usepackage{esvect}

\usepackage{ulem}

\usepackage{comment}

\usepackage[colorlinks=true, citecolor=blue, linkcolor=blue, urlcolor=blue]{hyperref}

\begin{document}
	
	\title{High-harmonic generation from artificially stacked 2D crystals}
	
	\author{Christian Heide} 
	\email{cheide@stanford.edu}
	\thanks{These authors contributed equally.}
	\affiliation{Stanford PULSE Institute, SLAC National Accelerator Laboratory, Menlo Park, CA 94025, USA}
	\affiliation{Department of Applied Physics, Stanford University, Stanford, CA 94305, USA}
	\author{Yuki Kobayashi}
	\thanks{These authors contributed equally.}
	\affiliation{Stanford PULSE Institute, SLAC National Accelerator Laboratory, Menlo Park, CA 94025, USA}
	\affiliation{Department of Applied Physics, Stanford University, Stanford, CA 94305, USA}
	\author{Amalya C. Johnson}
	\affiliation{Department of Materials Science and Engineering, Stanford University, Stanford, CA 94305, USA}
	\author{Tony F. Heinz}
	\affiliation{Stanford PULSE Institute, SLAC National Accelerator Laboratory, Menlo Park, CA 94025, USA}
	\affiliation{Department of Applied Physics, Stanford University, Stanford, CA 94305, USA}
	\author{David A. Reis}
	\affiliation{Stanford PULSE Institute, SLAC National Accelerator Laboratory, Menlo Park, CA 94025, USA}
	\affiliation{Department of Applied Physics, Stanford University, Stanford, CA 94305, USA}
	\author{Fang Liu}
	\affiliation{Stanford PULSE Institute, SLAC National Accelerator Laboratory, Menlo Park, CA 94025, USA}
	\affiliation{Department of Chemistry, Stanford University, Stanford, CA 94305, USA}
	\author{Shambhu Ghimire}
	\affiliation{Stanford PULSE Institute, SLAC National Accelerator Laboratory, Menlo Park, CA 94025, USA}
	\date{\today}

\begin{abstract}
	{We report a coherent layer-by-layer build-up of high-order harmonic generation (HHG) in artificially stacked transition metal dichalcogenides (TMDC) crystals in their various stacking configurations. In the experiments, millimeter-sized single crystalline monolayers are synthesized using the gold foil-exfoliation method, followed by artificially stacking on a transparent substrate. High-order harmonics up to the 19th order are generated by the interaction with an ultrafast mid-infrared (MIR) driving laser. We find that the generation is sensitive to both the number of layers and their relative orientation. For AAAA stacking configuration, both odd- and even-orders exhibit a quadratic increase in intensity as a function of the number of layers, which is a signature of constructive interference of high-harmonic emission from successive layers. Particularly, we observe some deviations from this scaling at photon energies above the bandgap, which is explained by self-absorption effects. For AB and ABAB stacking, even-order harmonics remain below the detection level, consistent with the presence of inversion symmetry. Our study confirms the capability of producing non-perturbative high-order harmonics from stacked layered materials subjected to intense MIR fields without damaging samples. It has implications for optimizing solid-state HHG sources at the nano-scale and developing high-harmonics as an ultrafast probe of artificially stacked layered materials. Because the HHG process is a strong-field driven process, it has the potential to probe high-momentum and energy states in the bandstructure combined with atomic-scale sensitivity in real space, making it an attractive probe of novel material structures such as the Moiré pattern.}
\end{abstract}%

\maketitle
\begin{figure}[t!]
		\begin{center}
			\includegraphics[width=8cm]{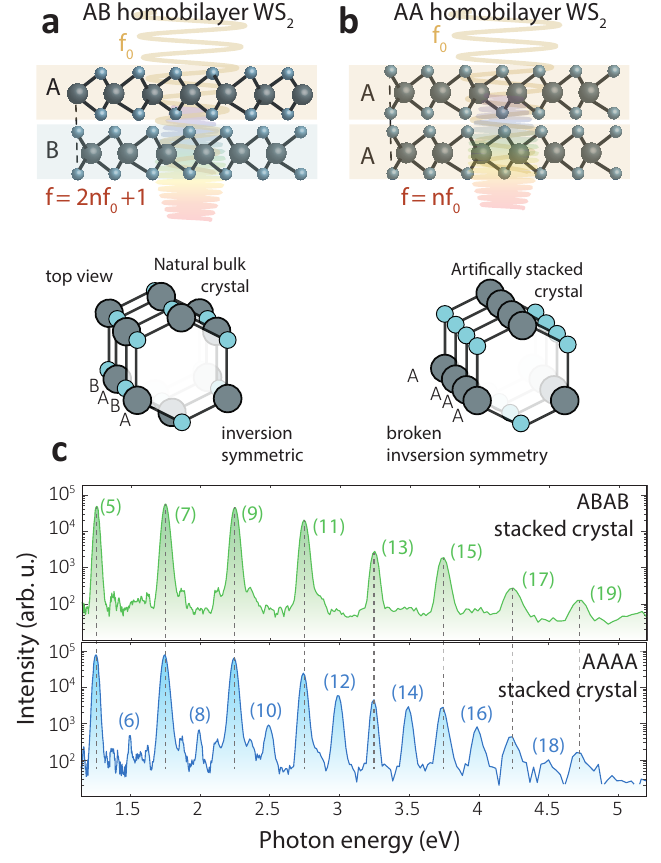}
			\caption{\textbf{High-harmonic generation from artificially stacked WS$_2$.}
				\textbf{a}, Schematic representation of high-harmonic generation in AA and AB stacked homobilayers of WS$_2$. \textbf{b}, Illustration of the underlying symmetry of naturally grown (AB) and artificially stacked (AA) WS$_2$. The natural bulk AB-stacked crystal is centrosymmetric, whereas the AA-stacked artificial crystal exhibits broken inversion symmetry, as highlighted by the dashed red line. \textbf{c}, Measured HH spectrum for 4-layered ABAB-stacked and AAAA-stacked TMDC crystals. The artificially stacked crystal with its broken inversion symmetry produces both even- and odd-order harmonics ranging from 5 to 19. }
			\label{fig: 1}
		\end{center}
	\end{figure}

{High-harmonic generation (HHG) is a strong-field optical process that has been well studied in atomic and molecular gases \cite{ferray_multiple-harmonic_1988, Rundquist1998, Lewenstein1994, Corkum2007}. In the phase-matched configuration, the harmonic signal adds coherently and the intensity increases quadratically as the gas density is increased. The quadratic scaling stems from constructive interference of the electric field of the high harmonics generated in each atom or molecule. More recently, driven by the development of long wavelengths laser sources, high-harmonic generation has been observed in solid-state samples \cite{ghimire2011observation}, which lead to a vibrant new research direction \cite{Vampa2015, Garg2016, Ghimire2019,Bai2020, Goulielmakis2022,Chang2022}. Research in the field has been directed towards an advanced understanding of the microscopic mechanism of strongly driven electrons in solids \cite{Vampa2015, Hohenleutner2015, Vampa2016, Garg2016, Liu2016, Yoshikawa2019, Uzan2020, Costello2021, Yue2022, UzanNarovlansky2022, Freudenstein2022, Yue2022}, characterization of the temporal profiles of the harmonics for the purpose of generation of atoosecond pulse \cite{You2016,Langer2017, Nourbakhsh2021}, probing electron-hole coherence in driven solid materials \cite{Heide2022}, probing light-field driven electronics \cite{Schiffrin2012, Higuchi2017, Heide2018, Neufeld2021, Heide2021Nanoletter, Boolakee2022}, in-situ focusing and shaping of high-harmonics using structured samples \cite{Sivis2017, Korobenko2021}, and most recently in probing the topological properties\cite{Bai2020, Schmid2021, Baykusheva2021, Heide2022b}. Due to the high density of the target, the treatment of macroscopic effects in solid materials requires careful consideration of absorption \cite{Liu2020b}, as well as self-reaction effects, such as self-phase modulation of the driving laser pulse \cite{Lu2019}. Often the desired thickness is in the nanoscale \cite{Ghimire2012}, limited by either phase mismatch or absorption.} \\
Transition metal dichalcogenides (TMDC) are layered van-der-Waals materials that provide an exceptional platform for material synthesis at the nanoscale. Contrary to their bulk counterpart, monolayers exhibit broken inversion symmetry, therefore second-harmonic generation (SHG) has been an interesting approach to studying electronic and optical properties as a function of the number of layers {and crystal orientation} \cite{Kumar2013, Li2013, Janisch2014, Hsu2014, Zhao2016, Shan2018, Liu2020}. Previously, non-perturbative HHG was observed from monolayer TMDCs subjected to intense mid-infrared laser fields \cite{Liu2016, Yoshikawa2019}. Unlike their bulk counterpart, monolayers exhibit broken inversion symmetry and, correspondingly, produce even-order harmonics in addition to the odd orders. In natural bulk crystals the crystallographic orientation of adjacent layers is rotated by 180$^\circ$, which is denoted as AB (or Bernal) stacking, as schematically illustrated in Fig. \ref{fig: 1}\,\textbf{a} \cite{Hsu2014, Shinde2018, Shan2018, Liu2020, Yao2021}. The availability of individual monolayers from exfoliation and the possibility of their arbitrary stacking represent a material synthesis-based approach to control optical nonlinearities and thus to optimize solid-state HHG, which recently gained tremendous attention due to the observation of superconducting states in graphene moiré superlattices \cite{Cao2018} and the observation of flat bands in a wider range of twist angles using TMDC \cite{Zhang2020}. This artificial stacking allows one to study how the efficiency for the generation of even- and odd-order harmonics evolves as a function of a number of layers, including the influence of macroscopic effects such as absorption and phase mismatch, as well as potential interlayer coupling effects.

	\begin{figure*}[t!]
		\begin{center}
			\includegraphics[width=16cm]{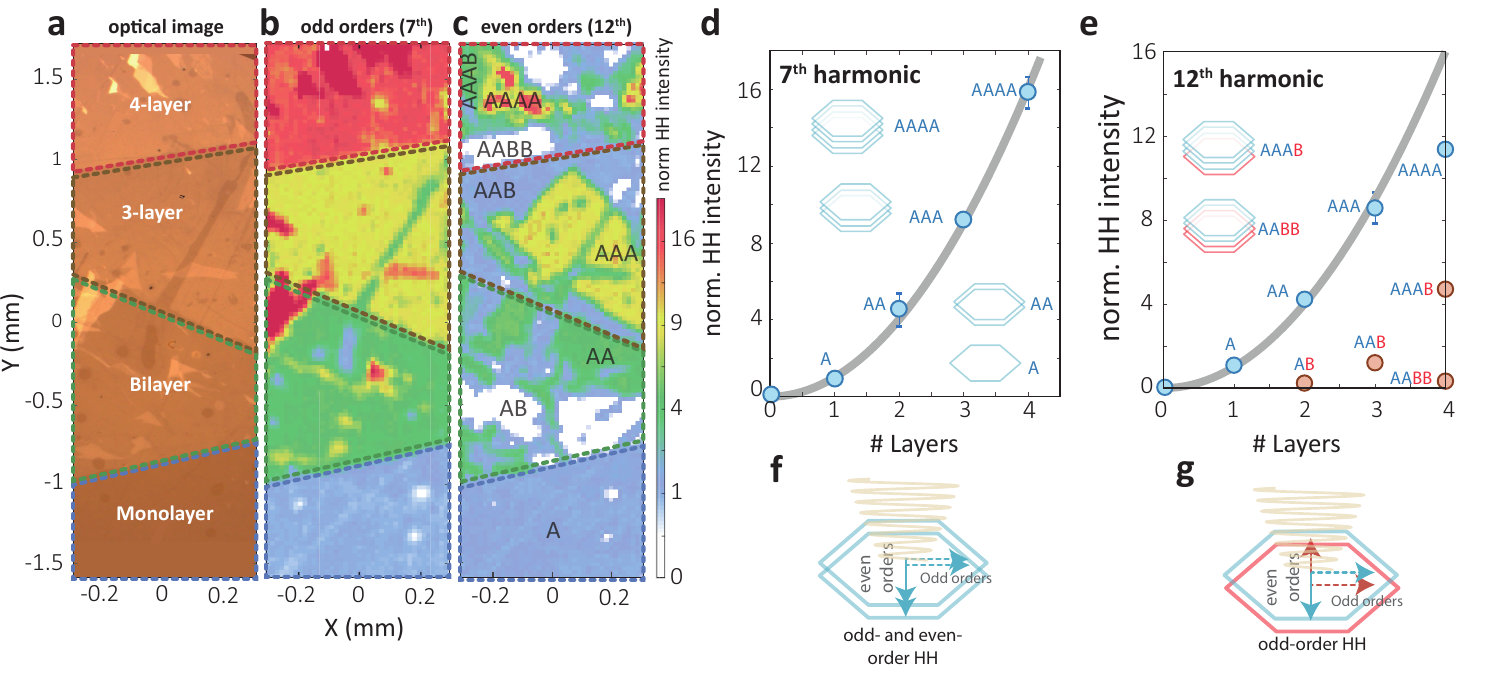}
			\caption{\textbf{Spatial mapping of HHG from mm-sized artificially stacked WS$_2$ crystals.} 
				\textbf{a}, optical image of large-scale stacked WS$_2$. The indicated regions of different layer thickness have been identified by optical contrast measurements. 
				\textbf{b, c}, HH mapping for HO 7 and HO 12, as representative even and odd-order HH signals. \textbf{d}, For A...A stacking, we find quadratic scaling (gray line) of HO 7 as a function of layer thickness. The mean value and the error bar of the data points are obtained by averaging over the stacked area, normalized to the the monolayer. {Number of layers equal to zero corresponds to the condiation when only the substrate is illuminated.} \textbf{e}, Similarly, quadratic scaling of HO 12 as a function of layer thickness is found, with some deviation for samples with four layers. As a result of the sensitivity to inversion symmetry, the even-order harmonic response is distinct for AA and AB stacking. \textbf{f, g}, Illustration of even and odd-order harmonic generation for AA stacking in (f) and AB in (g), respectively.}
			\label{fig: 2}
		\end{center}
	\end{figure*}

	Here, we report first results on high-harmonic generation in artificially stacked crystals as a function of the number of layers and their stacking domain. We observe that the HHG response is dominated by constructive interference from adjacent layers, close to the ideal limit a coherent layer-by-layer build-up of high-order optical harmonics results in a quadratic optical response $I_\text{HHG} \propto n^2$, with $n$ the number of stacked layers. For harmonics above the bandgap we observe some deviations which are attributed to absorption.

Monolayers of tungsten disulfide (WS$_2$) are prepared using a novel gold-assisted exfoliation technique. This method enables the creation of mm-sized single-crystalline flakes, as shown in Fig. \ref{fig: 2}\,\textbf{a} \cite{Liu2020}. These large monolayers are then stacked on a fused silica substrate to form a tailor-made crystal with mm-sized in-plane dimensions and the desired symmetry properties. The angles between adjacent layers are aligned with a precision of about $\pm$5$^\circ$, limited by our fabrication technique. The artificial crystals are illuminated at normal incidence and ambient conditions with linearly polarized laser pulses with a pulse duration of $\sim$100~fs and at a central wavelength of $\sim$ 5~$\mu$m. The laser beam is focused on the sample with a spot size of $\sim$ 90~$\mu$m (1/$e^2$ intensity radius). For the laser pulse energy of 4~$\mu$J, this yields an estimated vacuum peak intensity of $\sim 3 \times$10$^{11}$~W/cm$^2$, corresponding to a peak fluence of $\sim 31$~mJ/cm$^2$ or a peak vacuum field strengths of $\sim 1.2$~V/nm at the sample. This peak intensity is below the damage threshold of the sample at the experimental 1\,kHz repetition rate. The generated high harmonics are collected from the sample in transmission and focused by CaF$_2$ lenses into a spectrometer. \\

	For crystals with AA stacking, the broken inversion symmetry of the monolayer is preserved, as illustrated in Fig.\,\ref{fig: 1}\,\textbf{b} \cite{Li2013, Malard2013}. For the case of second-harmonic generation (SHG), it has been shown that the response then scales quadratically with the number of stacked monolayers \cite{Hsu2014, Zhao2016, Shan2018, Liu2020}. Here we extend these measurements to the strong-field regime, where the harmonic spectrum extends up to harmonic order (HO) 19. Figure \ref{fig: 1}\textbf{c} shows the HH spectra for four layers of WS$_2$, for both ABAB (top panel) and AAAA (bottom panel) stacking. The spectra exhibit distinct peaks at integer multiples of the pump photon energy, corresponding to high harmonics ranging from the 5th to 19th order, which, on the lower photon energy side, is limited by the response of our detection system (see SI). For the ABAB-stacked samples, no even-order harmonics are detected. This is consistent with the inversion symmetry of the structure, a property that emerges from pairs of AB stacked layers (and is also present in the bulk material). In contrast, for AAAA stacking, both even- and odd-order harmonics are observed. The in-plane sample orientation in these measurements was aligned to maximize even-order harmonics for the linearly polarized MIR laser field, which corresponds to the pump electric field lying along the $\Gamma-K$ direction of the crystal thickness. \cite{Liu2016}. 
	
	We next discuss how the intensity of each harmonic order depends on the number of stacked layers. We fabricated samples that consisted of one to four stacked layers, as shown in Fig.\,\ref{fig: 2}\,\textbf{a}. Based on the optical contrast, we find four regions in our large sample: from the bottom to the top, a monolayer region, followed by a bilayer, a trilayer, and finally a four-layer WS$_2$ crystal. Since the spot size of the laser is much smaller than the lateral size of the crystal, we can collect HHG spectra as a function of position across the sample. By integrating over the energy for each harmonic order, we can generate high-harmonic maps, as shown in the SI. Figure\,\ref{fig: 2}\,\textbf{b} presents such a map for the 7th harmonic order, which we use as a representative of the odd-order response. The intensity is normalized to the intensity of the monolayer. Here we also observe four different regions, in agreement with the contrast optical image. The normalized intensity is plotted as a function of layer thickness in Fig.\,\ref{fig: 2}\,\textbf{d}. We find that for two layers we obtain 4 times the HH intensity compared with the monolayer, for 3 layers 9 times, and for 4 layers 16 times, i.e., a quadratic scaling of the harmonic yield with the number of stacked layers. 
	
	We also analyze the high-harmonic map of the 12th harmonic order as a representative of even-order harmonics for the different regions of the sample. First, we note that the even-order response is predominantly polarized perpendicular to the electric field of the mid-IR pump, which can be attributed to the valley-contrasting Berry curvature resulting from the broken inversion symmetry in the monolayers \cite{Liu2016, Kobayashi2021}. The Berry curvature gives rise to an anomalous Hall contribution, resulting in even-order harmonics polarized normally to the pump field. The phase of the even-order harmonics is determined by the sign of the Berry curvature. Rotating one layer by 180$^\circ$ flips the sign of the Berry contribution and, thus, the anomalous Hall contribution of this layer is phase shifted by $\pi$. This phase shift causes destructive interference of even-order harmonics in the case of an AB-stacked sample, as illustrated schematically in Fig.\,\ref{fig: 2}\,\textbf{g}. In case of the AA-stacked sample, the contributions of the even-order contribution from both layers are in phase and interfere constructively (see \ref{fig: 2}\,\textbf{f})\\
	This stacking-dependent interference is observed in the even-order HH intensity map shown in Fig.\,\ref{fig: 2}\,\textbf{c}. In contrast to the odd-order harmonic response, we find several sub-domains in samples of a given layer thickness. For example, we observe in the bilayer and 4-layer region areas where no even-order harmonic emission is present, despite efficient odd-order harmonic generation. Plotting the normalized HH intensity allows us to assign these domains to different stacking configurations, as indicated in Fig.\,\ref{fig: 2}\,\textbf{e}. For example, the trilayer region labeled as AAB-stacking matches the intensity of a monolayer, and the 4-layer AAAB region matches the intensity of a bilayer of AA-stacked material. We point out that our measurement is not able to distinguish between AAB stacking and permutations of this stacking order. 
	For pure AA stacking, we observe that, just as for the odd-order harmonics, even-order harmonics scale quadratically with the number of layers, with modest deviations for the four-layer samples. This analysis shows that one harmonic order is not sufficient to completely determine the stacking configuration. For example, one odd-order harmonic measurement to determine the total number of layers and one even-order harmonic measurement to determine the number of AB combinations would be required to distinguish a monolayer from an AAB-stacked trilayer or no sample from an AB-stacked bilayer. 
	We note that the domains with different A/B layer orientations most likely originate from the cleaving process that precedes the exfoliation and produces terrace steps with alternating domain orientation. 
	\begin{figure*}[t!]
		\begin{center}
			\includegraphics[width=16cm]{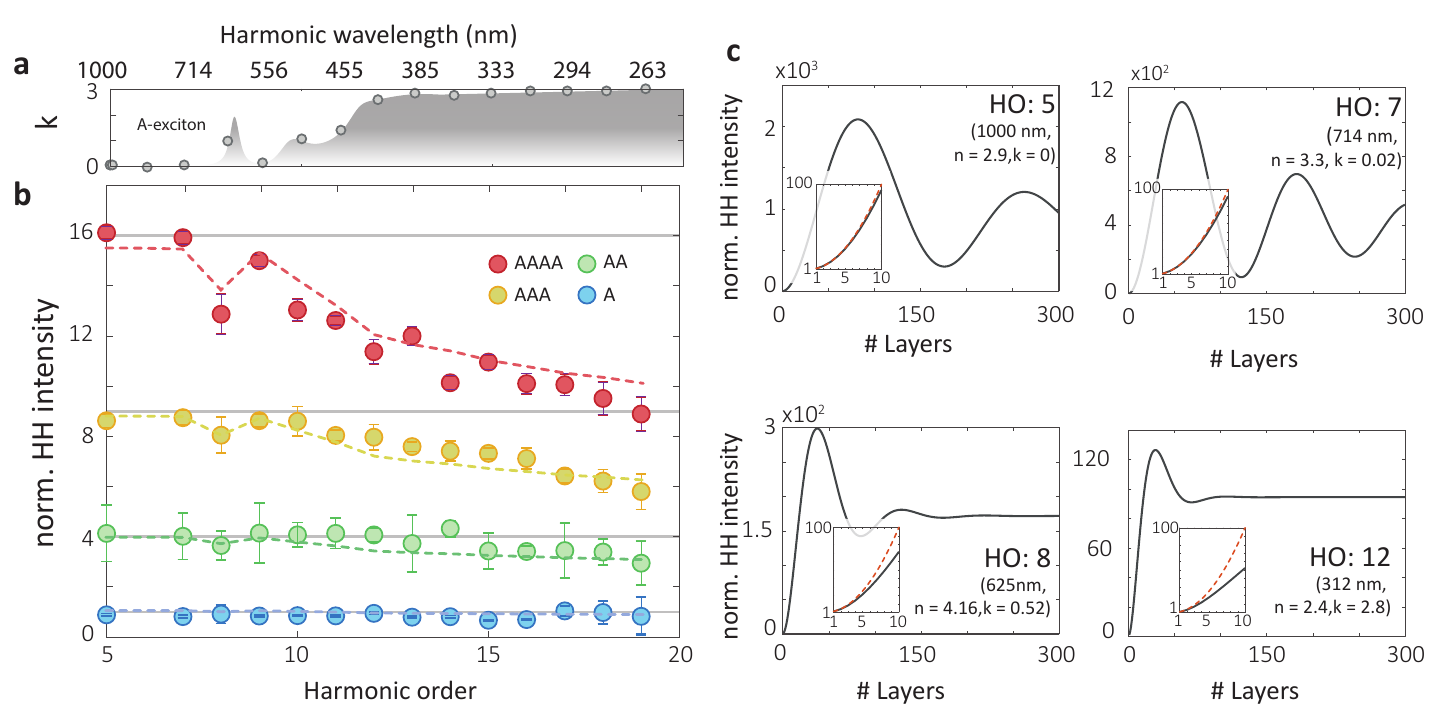}
			\caption{\textbf{High-harmonic generation from artificially stacked materials as a function of layer thickness.}
				\textbf{a} Imaginary part of the refractive index k of WS$_2$ for in-plane response \cite{Liu2020_refractive}.
				\textbf{b} Normalized HH intensity as a function of the harmonic order for one layer (A, blue dots), two layers (AA, green dots), three layers (AAA, orange dots), and four layers (AAAA, red dots). Deviations from a quadratic scaling as a function of AA-stacked layers are mainly found above HO 10, where the absorption in WS$_2$ becomes large. The dashed line is the calculated harmonic efficiency taking both phase-matching and re-absorption effects into account. \textbf{c}, The calculated HH intensity for four different harmonic orders as a function of crystal thickness based on both the real and imaginary refractive indices. Inset: Detail of the prediction from 1 to 10 layers. The red line is a reference quadratic variation.}
			\label{fig: 3}
		\end{center}
	\end{figure*}
	
	There are a number of physical effects in the material that cause the layers to not be fully independent, such as mutual dielectric screening and exciton binding energy changes, band hybridization effects, strain fields, and modulation of the interlayer spacing. Furthermore, absorption of the driving field and generated high harmonics as well as phase-matching may result in deviation from a ideal quadratic scaling. 
	
	To understand the deviations from a purely quadratic scaling for HO 12, we plot in Fig.\,\ref{fig: 3}\,\textbf{b} the normalized harmonic yield for all AA-stacking configurations as a function of the harmonic order. The mean and the error value are obtained by integrating over the HH counts of different stacking regions. For photon energies of harmonics above the bandgap, we note a deviation from perfect quadratic scaling. We compare the experimental results with expectations for the HHG process allowing for the relative phase mismatch and attenuation of the pump and different harmonic orders using the corresponding complex refractive index. This calculation thus takes into account the absorption the HH radiation is it propagates through the stacked sample, as well as the effect of phase mismatch, i.e., the interference of the HH generated from each layer. We see that the propagation effects capture the observed deviation from quadratic scaling quite well for the different harmonic orders (dashed lines). Simulations for up to 10 stacked layers for four different harmonic orders are displayed in the insets of Fig.\ref{fig: 3}\,\textbf{c} and in Fig.\,\ref{fig: 3}\,\textbf{b}, with the red curve showing quadratic scaling for comparison. {We note that compared to second and third harmonic generation, HH is also sensitive to the absorption coefficient in a broad spectral region.}
	
	For harmonics below the bandgap, i.e., 5th and 7th harmonics, the phase mismatch between the harmonic and the pump beam limits the efficiency, which becomes a prominent effect only for more than 30 layers. Above the bandgap ($\sim$ 2~eV, $\sim$220~nm) and close to the A-exciton resonances, the efficiency is mainly limited by absorption, as seen for the 8th and the 12th harmonic, respectively. While the propagation of the below bandgap harmonics is modulated mainly by phase mismatch and can reach harmonic enhancement factors of 100 for 10 stacked layers, the above bandgap harmonics are absorption limited. 
	
	In this letter, we have focused on homostructures artificially stacked in AB and AA configurations, with an angular alignment accuracy of about $5^\circ$. Our result indicate that despite various complexities in the stacked sample the predominant effects in the HHG spectrum can be captured by looking at non-interacting layers. However, a more precise stacking alignment of the homo- or heterostructures may allow the creation of atomically thin p-n and Schottky junctions \cite{Yang2019, Heide2020} and twisted monolayers with moiré lattices \cite{Balents2020, Barr2022}, which may exhibit superconductivity \cite{Cao2018}, magnetism \cite{Sharpe2019}, topological edge states \cite{Huang2018}, and other emergent properties. Since high-harmonic generation is sensitive to the crystal structure \cite{You2016} and the band structure \cite{Vampa2015}, it would be interesting to use this process as a {microscopic} probe of moiré systems. In particular, the electron trajectories in real space under high-field conditions may be tuned over distances comparable or longer than the moiré confinement potential. As an upper boundary we compare the maximum excursion distance of semi-classical electrons in a solid $r_\text{max}$ with the period of the moiré superlattice $a_\text{m}$. For example, for WS$_2$ homobilayers and a twist angle of 3$^\circ$ one obtain $a_\text{m} = $ 6\,nm \cite{Zhang2020}. $r_\text{max}$ can be estimated as $eE\lambda^2/4\pi^2mc^2$, where $E$ is the pump electric field strength, $\lambda$ the pump wavelengths, $e$ the elementary charge, and $m^\ast$ the reduced electron mass. For WS$_2$ ($m^\ast = 0.34\,m_0$ \cite{Ovchinnikov2014}) and, for the laser parameters in our experiment, we achieve $r_\text{max} = 4\,$nm, which is comparable to $a_\text{m}$ and may allow the observation of moiré effects in twisted homobilayers.
	
	As for future applications of compact high-harmonic generation sources to obtain short wavelengths emission, it may be beneficial to work with wide-bandgap monolayers, such as hexagonal boron nitride to avoid re-absorption effects in thicker samples. In this case, intensity enhancement factors of roughly three orders of magnitude would be expected before the phase mismatch limits the HHG efficiency. Furthermore, in the few-layer limit, the high-harmonic radiation can further be enhanced and shaped by meta-surfaces \cite{Vampa2017, Liu2018, Yang2019, Liu2020b, Jalil2022}, which may turn artificially stacked monolayers into an attractive high-harmonic source for pump-probe experiments with tunable even/odd high harmonics. 
	
	{In conclusion, we observed the coherent growth of high-harmonic generation from individual layers of artificially stacked monolayers of the van-der-Waals TMDC semiconductor WS$_2$.} HHG is sensitive to sample domains as well as to the number of layers and their various stacking configurations. We use mm-sized flakes that are much bigger than typical laser spots, which are on the order of 100 $\mu$m. On the materials side, unlike the conventional scotch tape exfoliation technique or chemical vapor deposition grown methods, which yield single-crystalline flakes smaller than 100 $\mu$m, here we use gold-assisted exfoliation to produce mm-sized single-crystalline TMDCs. On the optical side, we use ultra-short laser pulses in the mid-infrared wavelength range and therefore can produce a plateau of harmonics without damaging the sample. In addition, our results also represent an important step towards designing efficient and functional high-harmonic sources by artificially tailoring solid-state materials, with ramifications in light-field driven electronics. 
	
%

	\section{Acknowledgments}
	This work was supported by the US Department of Energy, Office of Science, Basic Energy Sciences, Chemical Sciences, Geosciences, and Biosciences Division through the AMOS program. F.L. acknowledges support from a Terman Fellowship and startup funds from the Department of Chemistry at Stanford University. Y.K. acknowledges support from the Urbanek-Chorodow Fellowship from Stanford University and C.H. from the Humboldt Fellowship and the W. M. Keck Foundation. 
\end{document}